\documentclass[9pt,twocolumn,twoside]{osajnl}
\journal{ao} % Choose journal (ao, aop, josaa, josab, ol, optica, pr)

\setboolean{shortarticle}{false}
\usepackage{float}
\usepackage{upgreek}
\usepackage{graphicx}
\usepackage{gensymb}
\usepackage{hyperref}
\usepackage{siunitx}
\usepackage[normalem]{ulem}
\title{A simple, powerful diode laser system for atomic physics}

\author[1,*]{Andrew Daffurn}
\author[2]{Rachel F.\ Offer}
\author[1]{Aidan S.\ Arnold}

\affil[1]{Dept.\ of Physics, SUPA, University of Strathclyde, %107 Rottenrow,
Glasgow, G4 0NG, UK\\}
\affil[2]{Institute for Photonics and Advanced Sensing (IPAS) and School of Physical Sciences, University of Adelaide, Adelaide, SA 5005, Australia}

\affil[*]{andrew.daffurn@strath.ac.uk}

\begin{abstract}
External-cavity diode lasers are %deservedly
ubiquitous in atomic physics and a wide variety of other scientific disciplines, due to their excellent affordability, coherence length and versatility. However, for higher power applications, the combination of seed lasers, injection-locking and amplifiers can rapidly become expensive and complex. Here we present a %useful, 
high-power, single-diode laser design with specifications: $>$\SI{210}{\milli \watt}, \SI{100}{\milli \second}-linewidth \SI[separate-uncertainty]{427 \pm 7}{\kilo \hertz}, $>$99\% mode purity, \SI{10}{\giga \hertz} mode-hop-free tuning range and \SI{12}{\nano \metre} coarse tuning. Simple methods are outlined to determine the spectral purity and linewidth with minimal additional infrastructure. The laser has sufficient power to collect $10^{10}$ $^{87}$Rb  atoms in a single-chamber vapour-loaded magneto-optical trap. With appropriate diodes and feedback, the system could be easily adapted to other atomic species and diode laser architectures. %%Last sentence?
\end{abstract}

\setboolean{displaycopyright}{true}

\begin{document}

\maketitle

\section{Introduction}

The invention of the laser \cite{maiman_stimulated_1960} was a landmark moment
for the scientific world and the ability to generate monochromatic, coherent light has opened avenues for research in many fields including atomic physics \cite{wieman_using_1991}, %khalighi_survey_2014,
telecommunications \cite{gibson_free-space_2004,kimble_quantum_2008}, and measurement science \cite{ludlow_optical_2015,elvin_cold-atom_2019}. Its use in atomic physics precipitated the fields of laser spectroscopy \cite{micke_coherent_2020, ponciano-ojeda_absorption_2021} and laser cooling, which have led to applications in frequency and timing metrology \cite{mcgrew_atomic_2018,arnold_blackbody_2018,takamoto_test_2020}, magnetometry \cite{budker_optical_2007,ingleby_vector_2018} and inertial sensing \cite{dutta_continuous_2016,bidel_absolute_2018,zhai_talbot-enhanced_2018,overstreet_effective_2018}.

External-cavity diode lasers (ECDLs) are a popular choice for these purposes  \cite{macadam_narrowband_1992,ricci_compact_1995,arnold_simple_1998,cook_high_2012,brekke_3d_2020} and they can be constructed by operating a semiconductor diode in conjunction with optical feedback provided by a diffraction grating. The diode's sensitivity to feedback tempers the undesirable spectral properties of the device \cite{lang_external_1980}, narrowing the linewidth by orders of magnitude (also by feedback-free routes \cite{mcdowall_acousto-optic_2009}). This generally results in a device with high tunability and narrow linewidth at a reasonable cost. The atomic or molecular species and transition of interest determines the choice of components, particularly the laser diode and grating, however a wide range of wavelengths are accessible from home-made ECDLs for assisting laser cooling of K at \SI{405}{\nano \metre} \cite{unnikrishnan_sub-doppler_2019} or Sr at \SI{497}{\nano \metre} \cite{schkolnik_extended-cavity_2019}, to compact ECDLs for water vapor absorption at \SI{1.4}{\micro \metre}  \cite{jimenez_narrow-line_2017}. 

The ever-growing power demands and complexity of atomic physics experiments have resulted in the development of various solutions to fulfil a multitude of requirements. Where great stability and reliability are required this can be achieved with, e.g.: lasers using interference-filter based feedback \cite{baillard_interference-filter-stabilized_2006,thompson_narrow_2012}; 
modular laser systems demonstrating month-long sub-MHz locking  \cite{sahagun_simple_2013}; micro-integrated ECDLs with no movable parts for harsh, challenging space-based environments \cite{luvsandamdin_development_2013}; and  distributed bragg reflector (DBR \cite{pino_miniature_2013}) or distributed feedback (DFB \cite{gaetano_sub-megahertz_2020}) laser systems. 

For applications requiring considerable power, one can conveniently pass a few mW of power from a narrow linewidth  seed diode laser system through a tapered amplifier (TA), preserving the seed linewidth but with output power of order \SI{1}{W} \cite{nyman_tapered-amplified_2006,kangara_design_2014}. Due to poor mode quality typically around half the TA power is lost in fibre coupling \cite{voigt_characterization_2001}. Commercial TAs can cost tens of thousands of pounds, although bespoke `home-made' systems can be made by skilled users for a fraction of the cost, albeit with more assembly time. 

High-power alternatives include frequency-doubling telecommunications wavelength light in a nonlinear crystal or waveguide \cite{sane_11_2012} and sum-frequency mixing.  Ti:Sapphire laser systems \cite{adams_precision_1992} have additional benefits in terms of linewidth and intensity stability, however there is a concomitant increase in cost into the £100k range.

Here, we present an intermediate system that retains the low cost, narrow linewidth advantage of typical ECDLs, but uses a newly available $300\,$mW high-power diode to bridge the power gap to TAs. Our design reduces the financial barrier for developing $\gtrsim$100$\,$\si{\milli \watt} experiments, a power range that is routinely required for alkali atom laser cooling systems.
Most alkali metal atoms have stretched-state transition \cite{stretchedstate} saturation intensities of order $\approx1\,$mW/cm$^2$. For single-beam intensities beyond this saturation intensity atom number also saturates, and hence the $210\,$mW output power we achieve is already sufficient to saturate e.g.\ a single-cell vapour-loaded $^{87}$Rb magneto-optical trap with $10^9$ \cite{aathesis} to $10^{10}$ \cite{zhai_talbot-enhanced_2018} atoms using total powers of $20-200\,$mW split into three orthogonal pairs of $2.5-5.0\,$cm diameter retro-reflected beams.

Our findings support the idea that modern medium power commercial diodes, using architectures like quantum well and ridge waveguide, still function well in ECDL setups. In addition, we characterise the spectral purity and linewidth of our high power ECDL using simple methods which require minimal additional diagnostic equipment. 

\section{ECDL}

Traditional double heterostructure laser diodes used in ECDLs are limited to output powers of 10s of milliWatts.  However, more complicated diode architectures are now available for relatively low cost.  The diode \cite{ThorlabsDiode} in our high power ECDL uses a combined quantum well and ridge waveguide structure, which allows $300\,$mW output power to be reached at a typical central wavelength of \SI{785}{\nano \metre}.  

To test this new diode we retrofitted it to a pre-existing Littrow configuration ECDL.  The device is based on an inexpensive and easily manufactured design \cite{arnold_simple_1998}, which includes only simple modifications to commercially available components. More information on the design and electronics are available in Appendix A and D of Ref.~\cite{aathesis}, and we note minor subsequent changes in a footnote \cite{laserchanges}.  
Feedback is provided by an $1800\,$lines$\,$mm$^{-1}$ visible reflective holographic grating \cite{ThorlabsGrating}. This produces \SI{11.4}{\percent} feedback at operational wavelengths with the light polarised parallel to the lines of the grating, balancing wavelength selectivity with output power. Higher output powers  should be achieved using gold-coated $(+15\%)$ and/or UV $(+12\%)$ gratings, although the latter will reduce feedback.

\begin{figure}[!b]
\includegraphics[width=0.49 \textwidth]{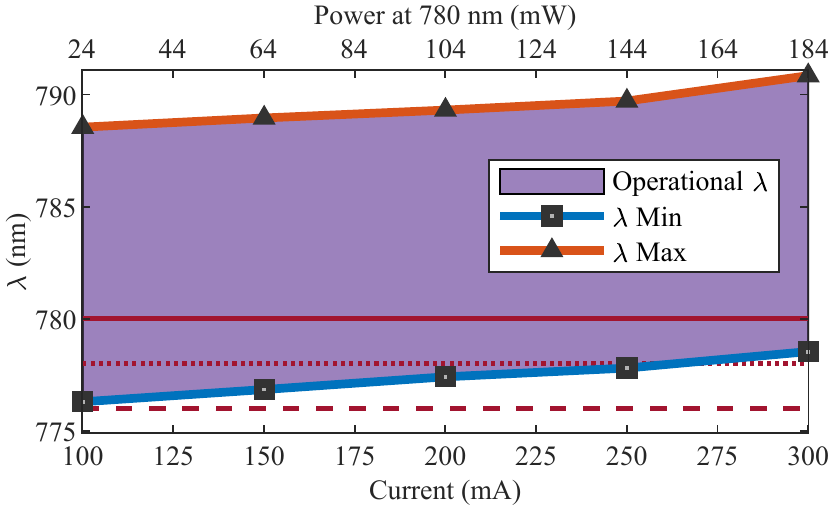}
\centering
\caption{Wavelength tuning range of the ECDL at \SI{14}{\celsius}. Also shown are the \SI{780}{\nano \metre} (solid line), \SI{776}{\nano \metre} (dashed line), and 2-photon \SI{778}{\nano \metre} (dotted line) wavelengths, representing transitions in Rb. Note that, in suitably designed systems, extreme temperatures can dramatically extend the tuning range \cite{tobias_low-temperature_2016}.}
\label{OW}
\end{figure}

\begin{figure*}[!t]
\includegraphics[width=\textwidth]{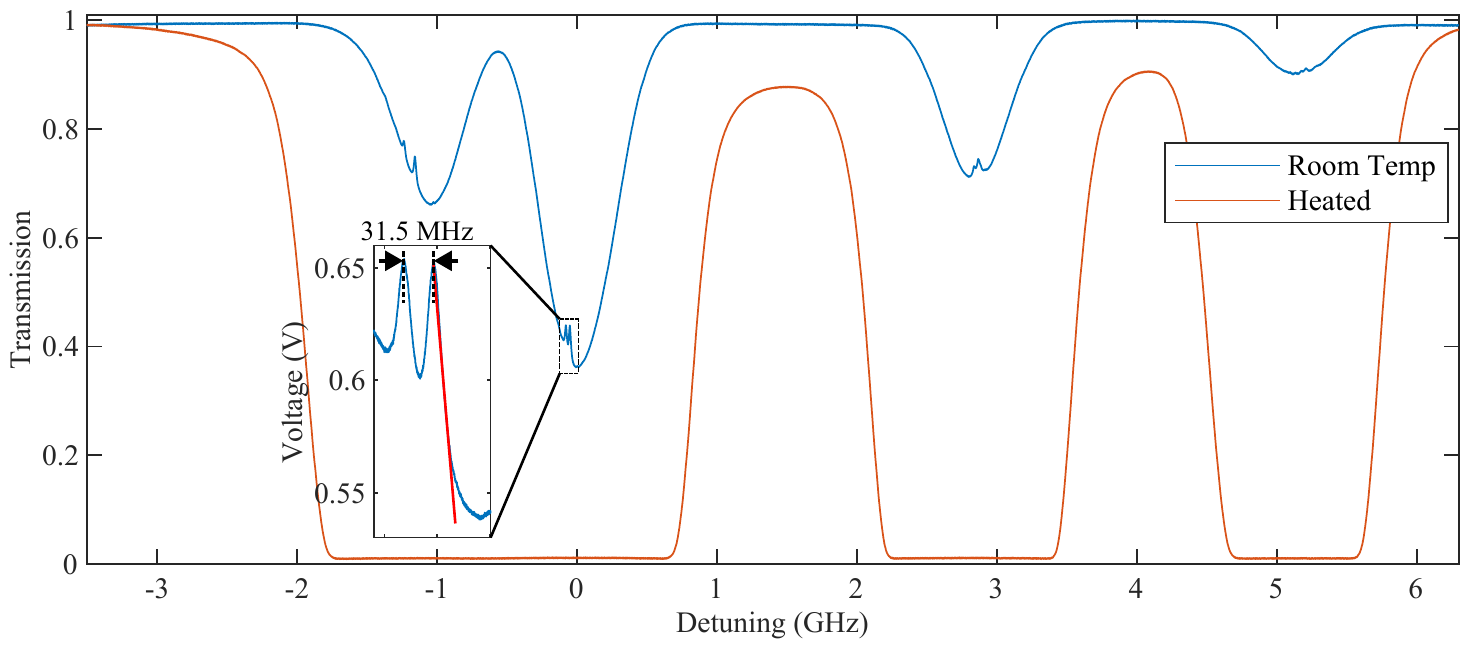}
\centering
\caption{Single-trace hyperfine pumping spectra of the Rb D$_2$ line at room temperature (blue) and in a heated cell (orange) with input beam intensity of \SI{17}{\milli \watt \per \centi \metre \squared}. The normalised photodiode absorption data have been adjusted to remove a \SI{-0.0083}{\per \giga \hertz} gradient introduced by the feed-forward scan. 
Inset: $^{85}$Rb $F=3\rightarrow F'=3,4$ crossover transition peak with red line to highlight slope where the free-running laser is `parked' for the linewidth measurement.}
\label{SatSpec}
\end{figure*}

With the high power diode installed, the ECDL achieves a \SI{780}{\nano \metre} output power of \SI{210}{\milli \watt} off the grating, before roll-off starts to occur at input currents of \SI{350}{\milli \ampere}.  We measure a diode lasing threshold of around \SI{75}{\milli \ampere}, and have demonstrated 70\% efficiency coupling to single mode fibre, indicating high spatial mode purity compared to a typical tapered amplifier. 

The diode was cooled with a \SI{33}{\watt} thermoelectric Peltier (TEC) to around \SI{14}{\celsius}. At this temperature it is always possible to address the D$_2$ Rb lines at \SI{780}{\nano \metre} at the operating currents for the device (Fig.~\ref{OW}). In humid environments with high dew point temperatures  \cite{Glasgow}, this represents the device's operational limit without compromising diode longevity, but the Rb 2-photon \SI{778}{\nano \metre} transition is already accessible (Fig.~\ref{OW}), and diode wavelength tuning is $\approx0.3\,\textrm{nm}/^{\circ}$C. By engineering the laser environment \cite{tobias_low-temperature_2016} dramatic temperature changes can vary available laser wavelengths by 10s of nanometres. 

\section{Mode-hop-free Range}

Ideally, a laser operates in a single resonator mode, which arises from the interplay between the semiconductor material, cavity length, external cavity, and feedback. An important characteristic is the mode-hop-free tuning of the device, this represents the maximum continuous frequency range the laser can scan before there is a modal jump. 
The material and cavity length are properties of the diode, and external feedback can be fine-tuned by altering the horizontal and vertical angle of the grating with respect to the diode. 
A short cavity is desirable to maximise the mode-hop-free scan range. However, a shorter cavity length has the effect of increasing the laser's linewidth. 
The external cavity length selected here of approximately \SI{20}{\milli \metre} allows for both a useful continuous scan range and narrow linewidth. 

For our device the mode-hop-free range was at least \SI{10}{\giga \hertz}, making it possible to continuously scan across all Doppler-broadened D$_2$ lines of $^{87}$Rb and $^{85}$Rb (Fig.~\ref{SatSpec}). This was achieved by adjusting the external cavity spacing using a piezo-electric transducer (PZT) and simultaneously scanning the current via a feed-forward signal \cite{doretff}. For this particular diode, scanning via the PZT alone yields a mode-hope-free range of only \SI{2}{\giga \hertz}.

\section{Amplified Spontaneous Emission}

 We are also interested in the percentage of coherent light produced by the laser, i.e.\ light from stimulated emission. Amplified spontaneous emission (ASE) that is reflected in the diode optical cavity produces lasing at threshold. However, an excess of ASE  limits the maximum gain in the material and contaminates the laser beam with a broad-spectrum incoherent background. 

 The mode purity of the light can be measured using an optical spectrum analyser, but these devices are expensive and generally have low resolution so are unable to spectrally resolve signals at the \SI{}{\mega \hertz} scale of the laser linewidth. A Fabry-Perot etalon can also be used, but this again necessitates additional equipment which requires careful alignment. We demonstrate an easy technique that simply requires a vapour cell -- which would already be required as part of a lab setup for locking the laser to an atomic line.

To measure the ASE of our device hyperfine pumping spectroscopy \cite{smith_role_2004}, widely misleadingly known as `saturated absorption spectroscopy', was performed on a heated Rb cell. Heating the cell dramatically increases the Rb atom number density in the cell, with an additional relatively minor effect on the width ($\propto \sqrt{T}$) of the Doppler-broadened absorption features \cite{siddons_absolute_2008,zentile_elecsus_2015,keaveney_elecsus_2018}.  For a completely coherent laser in a well-heated cell  \SI{100}{\percent} of the light would be absorbed at these Doppler broadened features, and any remaining light is a product of ASE. From Fig.~\ref{SatSpec} the remaining broadband ASE light that is transmitted through the cell is around \SI{1}{\percent}, matching what we have seen from other lower power \SI{780}{\nano \meter} laser diode systems.

\begin{figure}[!b]
\includegraphics[width=0.49\textwidth]{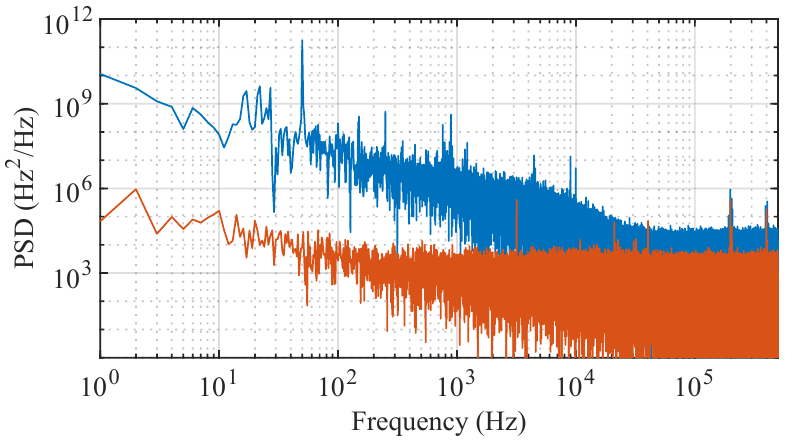}
\centering
\caption{Power spectral density from analysis of the noise data. Laser noise (blue) and background intensity noise (red). Analysis of the laser noise led to a $100\,$ms linewidth of \SI[separate-uncertainty]{427 \pm 7}{\kilo \hertz}.}
\label{PSD}
\end{figure}

\section{Linewidth}

Finally, we demonstrate a quick and easy method to measure the linewidth of a laser. A standard technique is to measure a radio frequency (RF) beat note, using a fast photodiode and an RF spectrum analyser \cite{Scholten2002, VonBandel:16}. However, a beat note requires at least two lasers (two technically only suffice if the laser linewidth to be measured is much larger than the reference laser). Alternatively a beat note can also be measured using a self-heterodyning method {\cite{Kaivola1998}}, but this requires a sufficiently long length of fibre for a given linewidth. The method described here only requires a vapour cell hyperfine pumping spectroscopy setup, which is already required for a sub-Doppler atomic lock.

We measure the linewidth by using a high resolution feature in the photodiode absorption spectrum as a frequency discriminator \cite{Kane1995}.  
The peak selected was the $^{85}$Rb D$_2$ 5S$_{1/2}$ $F=3\rightarrow 5$P$_{3/2}$ $F'=2,3$ crossover transition (Fig.~\ref{SatSpec} inset), because of the relatively large and linear peak slope (red line with magnitude of slope \SI{6.3}{\milli \volt \per \mega \hertz}), however in principle any peak could be selected. The calibration trace is recorded first, then the free-running laser is frequency-tuned to the side of the crossover pumping peak (red slope,  Fig.~\ref{SatSpec} inset) and the transmitted power fluctuations are recorded for $1\,$s at $1\,$MS/s on an oscilloscope, yielding inferred laser frequency vs.\ time.
This method yielded a root-mean-square (RMS) linewidth of \SI[separate-uncertainty]{427 \pm 7}{\kilo \hertz} for an averaging time of \SI{100}{\milli \second}, with an indicative standard error from 10 traces. A separate commercial diode laser had a free-running linewidth of \SI{510}{\kilo \hertz} using the same technique. 

The resulting power spectral density of one \SI{1}{\second} linewidth measurement is also displayed in Fig.~\ref{PSD}. The upper frequency end of the noise spectrum will be limited by the photodiode roll-off frequency of 8kHz \cite{freqRollOff}. Note the observed linewidth is nonetheless likely to be an upper estimate, as the \SI{50}{\hertz} mains peak contributes to half the linewidth, and we have also not added the complexity of detecting (and subsequently removing) the contribution from laser intensity noise.

To evaluate the accuracy of these linewidth measurements, we compared to a standard beat-note measurement technique, however a beat-note typically yields relative power as a function of frequency. The time-dependent photodiode traces from the two lasers were therefore also converted to histograms of relative power vs.\ laser frequency, to enable an approximate comparison to the beat-note. The histograms were approximately normally distributed, with Gaussian fits yielding RMS linewidths for the home-built and commercial lasers of \SI{490}{\kilo \hertz} and \SI{530}{\kilo \hertz}, respectively, similar to the raw RMS values above. This gave a combined linewidth of \SI{720}{\kilo \hertz} when the two linewidths were added in quadrature. This value can be compared favourably to the \SI{700}{\kilo \hertz} RF beat note RMS linewidth of the two lasers from a spectrum analyser, using a Gaussian fit to the beat note over the same $100\,$ms averaging time.
%\rachel{Andy - can you find any references for using our histogram technique to compare with a beat-note?  I had a look but no luck.  In Ref 52 they go from frequency(time) to power(frequency) following C-D-A-B in Fig. 5 - which may be more correct.  Our method is probably ok only for a rough comparison(?), so maybe we should make that clear.  }
%This confirms that the technique we outline is broadly accurate relative to a standard RF beat-note measurement.

Additionally, this method can be used to evaluate the time-dependence of the linewidth \cite{VonBandel:16}. A \SI{1}{\second} noise measurement can be split into smaller samples and used to evaluate the RMS linewidths for varying averaging times, a figure illustrating the linewidth's time-dependence is included as part of the dataset \cite{dataset}.

\section{Conclusion}

We have developed an economical home-build ECDL solution that can produce hundreds of mW of continuous power with a free-running \SI{100}{\milli \second} linewidth of \SI[separate-uncertainty]{427 \pm 7}{\kilo \hertz}. It is a single-unit, inexpensive source of moderate power for atomic physics experiments involving rubidium and is currently operating as one of the pump lasers in a four-wave mixing experiment \cite{offer_spiral_2018,offer_gouy_2021}. 
We have also detailed cost- and time-effective techniques to determine various useful laser characteristics, including the spectral purity and linewidth. 

For applications requiring sub-kHz linewidths, the laser system could  also work in conjunction with an appropriate high-finesse cavity lock \cite{legaie_sub-kilohertz_2018,moriya_sub-khz-linewidth_2020}.

\begin{backmatter}
\bmsection{Funding}
Funding via the Leverhulme Trust (Grant No.\ RPG-2013-386) and EPSRC (Grant No.\ EP/M506643/1).

\bmsection{Acknowledgments}
In addition to the support provided by the funding organisations, we are grateful for valuable discussions with Jonathan  Pritchard and Erling Riis.

%We are grateful for valuable discussions with Jonathan  Pritchard and Erling Riis as well as funding via the Leverhulme Trust (Grant No.\ RPG-2013-386) and EPSRC (Grant No.\ EP/M506643/1).

\bmsection{Disclosures}
The authors declare no conflicts of interest.

\bmsection{Data availability} Data underlying the results presented in this paper are available in Ref. \cite{dataset}.

\end{backmatter}
\bibliography{Refs.bib}

\begin{thebibliography}{10}
\bibitem{laserchanges}
{The piezo-electric transducer (PZT) is replaced with a (safer!) low-voltage
  stack in a sapphire/brass housing under a mirror-mount actuator, allowing PZT
  frequency scan directly using standard $\pm 15\,$V analogue locking circuits,
  with the $-15\,$V output connected to the (electrically isolated) PZT
  `ground' lead. A useful modification to the lock electronics is to include a
  digital switch allowing a controllable disconnection between the integrator
  and lock-in. Due to the low free-running laser linewidth, this allows the
  laser to be unlocked, jumped to other frequencies, jumped back, and relocked
  for excursions up to $\sim 1\,$s. Note the ICL8038 signal generator chip is
  obsolete, but alternatives exist.}
\bibitem{freqRollOff}
{The roll-off frequency is limited by the \SI{2}{\meter} RG58 coaxial cable to
  the oscilloscope, with estimated capacitance \SI{165}{\pico \farad} and a
  response time of $\approx$ \SI{20}{\micro \second} in parallel with the
  \SI{100}{\kilo \ohm} resistor of the photodiode circuit}.
\end{thebibliography}
\end{document}